\documentclass[longauth]{aa} 
\usepackage{graphicx}
\usepackage{natbib}
\usepackage{color}
\usepackage{txfonts}
\bibpunct{(}{)}{;}{a}{}{,} 
\begin{document}

 \title{Detailed spectral and morphological analysis of the shell type SNR RCW 86}


\author{H.E.S.S. Collaboration
\and A.~Abramowski \inst{1}
\and F.~Aharonian \inst{2,3,4}
\and F.~Ait Benkhali \inst{2}
\and A.G.~Akhperjanian \inst{5,4}
\and E.O.~Ang\"uner \inst{6}
\and M.~Backes \inst{7}
\and A.~Balzer \inst{8}
\and Y.~Becherini \inst{9}
\and J.~Becker Tjus \inst{10}
% ...
}
\institute{
Universit\"at Hamburg, Institut f\"ur Experimentalphysik, Luruper Chaussee 149, D 22761 Hamburg, Germany \and
Max-Planck-Institut f\"ur Kernphysik, P.O. Box 103980, D 69029 Heidelberg, Germany \and
Dublin Institute for Advanced Studies, 31 Fitzwilliam Place, Dublin 2, Ireland \and
National Academy of Sciences of the Republic of Armenia,  Marshall Baghramian Avenue, 24, 0019 Yerevan, Republic of Armenia  \and
Yerevan Physics Institute, 2 Alikhanian Brothers St., 375036 Yerevan, Armenia \and
Institut f\"ur Physik, Humboldt-Universit\"at zu Berlin, Newtonstr. 15, D 12489 Berlin, Germany \and
University of Namibia, Department of Physics, Private Bag 13301, Windhoek, Namibia \and
GRAPPA, Anton Pannekoek Institute for Astronomy, University of Amsterdam,  Science Park 904, 1098 XH Amsterdam, The Netherlands \and
Department of Physics and Electrical Engineering, Linnaeus University,  351 95 V\"axj\"o, Sweden \and
Institut f\"ur Theoretische Physik, Lehrstuhl IV: Weltraum und Astrophysik, Ruhr-Universit\"at Bochum, D 44780 Bochum, Germany
% ...
}

   \date{Accepted November 11, 2015}

\abstract
{}
{We aim for an understanding of the morphological and spectral properties of the supernova remnant RCW~86 and for insights into the production mechanism leading to the RCW~86 very high-energy $\gamma$-ray emission. } 
{We analyzed High Energy Spectroscopic\ System (H.E.S.S.) data that had increased sensitivity compared to the observations presented in the RCW~86 H.E.S.S. discovery publication. Studies of the morphological correlation between the 0.5 -- 1~keV X-ray band, the 2 -- 5~keV X-ray band, radio, and $\gamma$-ray emissions have been performed as well as broadband modeling of the spectral energy distribution with two different emission models.  
}
 { We present the first conclusive evidence that the TeV $\gamma$-ray emission region is shell-like based on our morphological studies. The comparison with 2 -- 5~keV X-ray data reveals a correlation with the 0.4 -- 50~TeV $\gamma$-ray emission. 
  The spectrum of RCW~86 is best described by a power law with an exponential cutoff at   $E_{cut} = (3.5 \pm  1.2_{stat})\  \mathrm{TeV}$ and a spectral index of $\Gamma \approx 1.6 \pm 0.2$. 
   A static leptonic one-zone model adequately describes the measured spectral energy distribution of RCW~86, with the resultant total 
   kinetic energy of the electrons above $1\ \mathrm{GeV}$ being equivalent to $\sim$0.1\% of the initial kinetic energy of a Type I a supernova  explosion ($10^{51}\ \mathrm{erg}$).
   When using a hadronic model, a magnetic field of $B \approx 100\ \mu\mathrm{G}$ is needed to represent the measured data.
   Although this is comparable to formerly published estimates, a standard E$^{-2}$ spectrum for the proton distribution cannot describe the $\gamma$-ray data. Instead, a spectral index of $\Gamma_p \approx 1.7$ would be required, 
 which implies that $\sim 7 \times 10^{49} / n_{\mathrm{cm}^{-3}} \mathrm{erg}$ has been transferred into high-energy protons with the effective density $n_{\mathrm{cm}^{-3}} =n/ 1 \, \mathrm{cm}^{-3}$.  This is about 10\% of the kinetic energy of a typical  Type Ia supernova under the assumption of a density of 1~cm$^{-3}$.}   
{}
 
\titlerunning{Detailed  analysis of the shell type SNR RCW 86}
\maketitle 



\section{Introduction}
\label{Sec::intorduction}
Supernova remnants (SNR) are prime candidates to be sources of  Galactic cosmic rays. The detection of very-high-energy $\gamma$-ray (VHE; E$>$100~GeV)  and nonthermal X-ray emission from SNRs has shown that they do in fact accelerate particles to energies above 100~TeV \citep[see, e.g.,][]{Hinton::2009,Aharonian::2013}. 

RCW~86 (also known as G315.4-2.3 or MSH 14-{\it 63}) is located at a distance of ($2.5 \pm 0.5$)~kpc \citep{helder::2013}. It is almost circular in shape with a diameter of about 40' clearly showing  a shell-like structure in the optical \citep{smith::1997}, radio \citep{kesteven::1987}, infrared \citep{williams::2011} and X-ray \citep{pisarski::1984} regimes. 
The source has been detected in $\gamma$-rays \citep{aharonian::2009,yuan::2014}, but a shell-like structure was not resolved.  RCW~86 is a young SNR, which is  about 1800 years old,        based on the probable connection to the historical supernova SN~185 recorded by Chinese astronomers in 185 AD \citep{stephenson::2002,smith::1997,dickel::2001,vink::2006}. The nature of the RCW~86 supernova (SN) progenitor was under intense discussion in the context of its young age and its unusually large size with a radius of $R \approx 15 d_{2.5}$~pc  ($d_{2.5}$: the distance to RCW~86 in units of $2.5$~kpc).
\citet{williams::2011} comprehensively studied all arguments about the type of the progenitor of RCW~86 and conducted hydrodynamic simulations to explain the observed characteristics. They concluded that RCW~86 is likely the remnant of a Type Ia supernova. The explosion was off-center  in a low-density cavity carved by the progenitor system \citep[see also][]{vink::1997,broersen::2014}. 
 
The physical conditions vary within the volume of RCW~86.  While  slow shocks have been measured \citep[$\sim$600 -- 800\,$\mathrm{ km s}^{-1}$;] []{long::1990,ghavamian::2001} in the
southwest (SW) and northwest (NW) regions along with relatively high post-shock gas densities  \citep[$\sim$2~cm$^{-3}$;][]{williams::2011}, X-ray measurements by \citet{helder::2009} indicate high velocities of $6000 \pm 2800 \,\mathrm{ km s}^{-1}$  in the northeast (NE) whereas optical measurements of specific knots in this region by \citet{helder::2013} show large variations in proper motions between $700$ -- 2000\,$\mathrm{ km s}^{-1}$ and low densities of $\sim$0.1 -- 0.3~cm$^{-3}$ \citep[see][]{yamaguchi::2011}. In the SW, NE, and to a lesser extent in the NW region, nonthermal X-ray emission is detected \citep{bamba::2000,vink::1997,vink::2006,williams::2011}. 
Near a region of Fe-K-line emission in the SW, there is nonthermal emission possibly synchrotron radiation from electrons accelerated at the reverse shock \citep{rho::2002}.

Here, we present a new analysis of very high-energy $\gamma$-ray data of RCW 86 taken with the 
High Energy Stereoscopic System (H.E.S.S.).  When the discovery of RCW 86 was published by the
 H.E.S.S. collaboration, the morphology in the TeV $\gamma-$ray regime could not be resolved owing to
  limited statistics \citep{aharonian::2009}. The new analysis benefits from significantly improved statistics
   and addresses whether RCW 86 has a shell structure in the VHE  $\gamma$-ray regime.  In addition,
    the larger data set allows for a broadband modeling of the spectral energy distribution  with both a
     leptonic and an hadronic one-zone model.
We restricted ourselves to two simple models since it is unreasonable to 
use more intricate models with additional parameters that cannot be determined owing to limited statistics. 


\section{H.E.S.S. observations and analysis method}

The High Energy Stereoscopic System is located in the Khomas Highland of Namibia at a height of about 1800~m above sea level. In its first phase, the system consisted of four identical imaging Cherenkov telescopes, each with a mirror area of 107~m$^2$ and a large field of view of $5^{\circ}$ \citep{bernloehr::2003}. The data presented in this paper were taken during this phase. 

In 2009 the H.E.S.S. Collaboration announced the discovery of RCW~86 in the VHE regime \citep{aharonian::2009}.
The source morphology of RCW 86 was thoroughly studied, but the existing data did not suffice to settle the question of whether the VHE emission originates from a shell or from a spherical region. Between 2007 and 2011, observations of the neighboring source HESS\ J1458$-$608 \citep{delosreyes::2011} and the scan of the Galactic plane  have added additional observation time to the data set, which now amounts to $\sim 57$~h.
The  zenith angles of these observations range from 36$^{\circ}$ to 53$^{\circ}$ and have  a mean value of 40$^{\circ}$. 
The data were recorded with pointing offsets between $0.5^{\circ}$ and $2.5^{\circ}$, average $1.1^{\circ}$, from the RCW 86 position.
The sensitivity and  angular resolution were improved compared to the former publication \citep{aharonian::2009},  using an advanced  analysis method \citep[for further detail, see][]{denaurois::2009} instead of the formerly used standard Hillas-parameter-based analysis. {Standard }cuts were used for spectral analysis and {faint} source cuts for morphological analysis. The faint source cuts have an improved angular resolution at the expense of lower statistics and a higher energy threshold of about $100\ \mathrm{GeV}$ \citep[for the cut definition, see][]{denaurois::2009}.  
 The results we present are consistent with the results of a multivariate analysis technique \citep{ohm::2009}, using an independent calibration scheme.

\section{Results}
\label{sec::Results}

\begin{figure} 
\resizebox{\hsize}{!}{\includegraphics{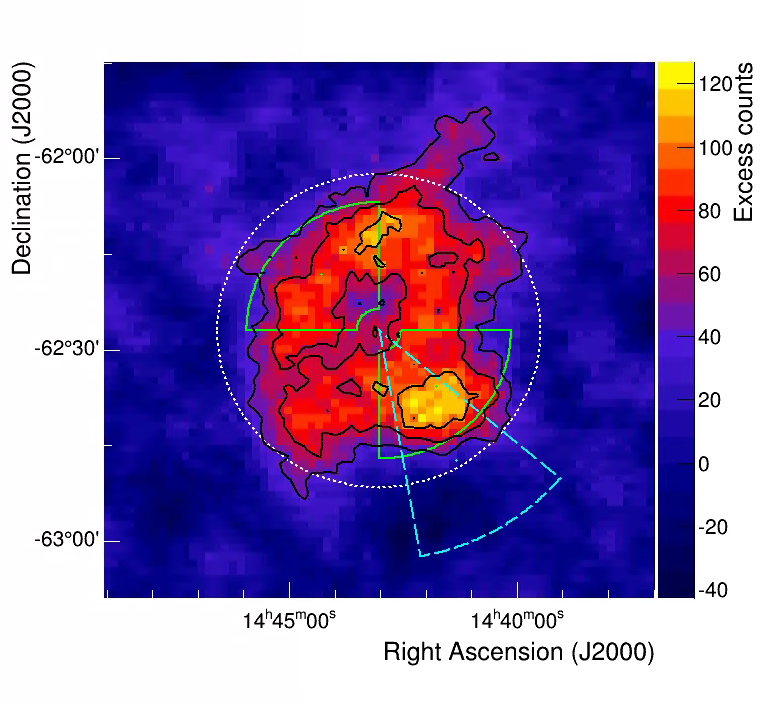}} 
\caption{VHE $\gamma-$ray emission of RCW~86. The sky map is extracted with faint analysis cuts and smoothed with a Gaussian filter with  $\sigma_{smooth} = $ 0\fdg06 to reduce the effect of statistical fluctuations. Black contours         correspond      to      3, 5, 7$\sigma$ significance. The white dotted circle depicts the integration region chosen for the spectral analysis. The dashed cyan sector shows the position angle range of the radial profile in Fig.~\ref{fig::RadialesProfile_rcw86}. The two green sectors (solid lines) give the extraction regions for spectra of the SW and NE regions discussed in Sect. \ref{sec::Results}. }
\label{fig::Skymap_rcw86}
\end{figure}

\begin{figure} 
\resizebox{\hsize}{!}{\includegraphics{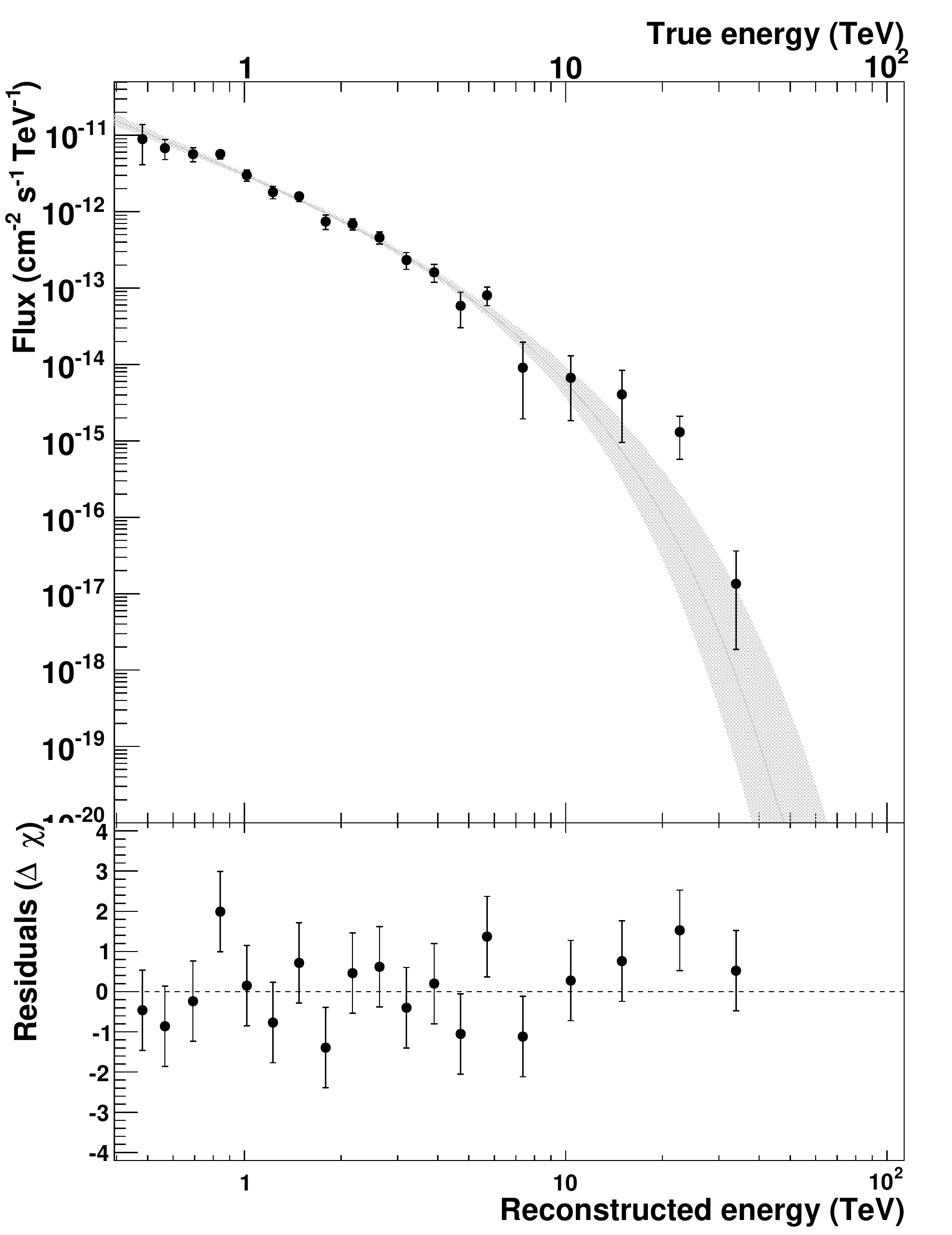}} \caption{In the upper panel, the differential energy spectrum of RCW~86 with the best-fit exponential cutoff power-law model is shown. The error bars denote 1$\sigma$ statistical errors. The shaded area represents the 1$\sigma$ confidence level of the fitted spectrum. In the bottom panel, the corresponding residuals are given.}
\label{fig::Spectrum_rcw86}
\end{figure}

 The $\gamma$-ray excess map, 
using the ring background subtraction technique \citep{berge::2007}, is shown in Fig. \ref{fig::Skymap_rcw86}. The map was smoothed with a Gaussian filter 
to reduce the effect of statistical fluctuations. The 68\% containment radius of  0\fdg061 of the H.E.S.S.  point spread function (PSF) 
was chosen as smoothing radius. 

Since this is larger than the bin size (0\fdg02), the pixels are correlated with each other. This is accounted for in the spatial-fitting process that we present. 
Extended $\gamma$-ray emission was found.   We detected $1220 \pm 87$ excess $\gamma$-rays within a  circular region centered at $\alpha_{J2000} = 14^{\mathrm{h}}43^{\mathrm{m}}2.16^{\mathrm{s}}$, $\delta_{J2000}= -62\degr26\arcmin56\arcsec $ with a radius of 0\fdg41  (see Fig.  \ref{fig::Skymap_rcw86}) and a source detection at a statistical significance of 18.3  $\sigma$.
The center is given by the flux averaged centroid, which was determined with the spectral results of  the previous H.E.S.S. publication of RCW~86 \citep{aharonian::2009}.  The radius was adjusted to the improved analysis method and  better angular resolution.

 \subsection{Spectral analysis}
A fit of a power-law function $\mathrm{d}N / \mathrm{d}E = \Phi_{0} (E/1 \mathrm{TeV})^{-\Gamma}$ to the spectral data analyzed with the standard analysis cuts yields $\Gamma  = 2.3 \pm  0.1_{stat} \pm  0.2_{sys}$,  and the differential flux normalization at 1~TeV of $\Phi_{0} =  (2.9 \pm 0.2_{stat} \pm  0.6_{sys}) \times 10^{-12}\mathrm{cm}^{-2}\mathrm{s}^{-1}\mathrm{TeV}^{-1}$ ($\mathrm{Log(likelihood)} = -54$; see Fig. \ref{fig::Spectrum_rcw86}). 
We tested if a power law with an exponential cutoff $\mathrm{d}N / \mathrm{d}E = \Phi_{0} (E/1 \mathrm{TeV})^{-\Gamma} \exp(-E/E_{cut})$  describes the data better than a simple power law. The fit gave  $\Phi(1\,\mathrm{TeV}) = (3.0 \pm 0.2_{stat} \pm  0.6_{sys}) \times 10^{-12}\mathrm{cm}^{-2}\mathrm{s}^{-1}\mathrm{TeV}^{-1}$ with $\Gamma  = 1.6 \pm  0.2_{stat} \pm  0.2_{sys}$  and a cutoff energy $E_{cut} = (3.5 \pm  1.2_{stat}\pm  2.2_{sys}) \mathrm{TeV}$ ($\mathrm{Log(likelihood)} = -47.7$ ), which is favored over the simple power law with a significance of 3.5 $\sigma,$ according to a log-likelihood ratio test. The spectral result of the power-law fit is compatible with a previous H.E.S.S. publication \citep{aharonian::2009}.

 \subsection{Morphology}
\begin{figure*}
\centering
\includegraphics[width=17cm]{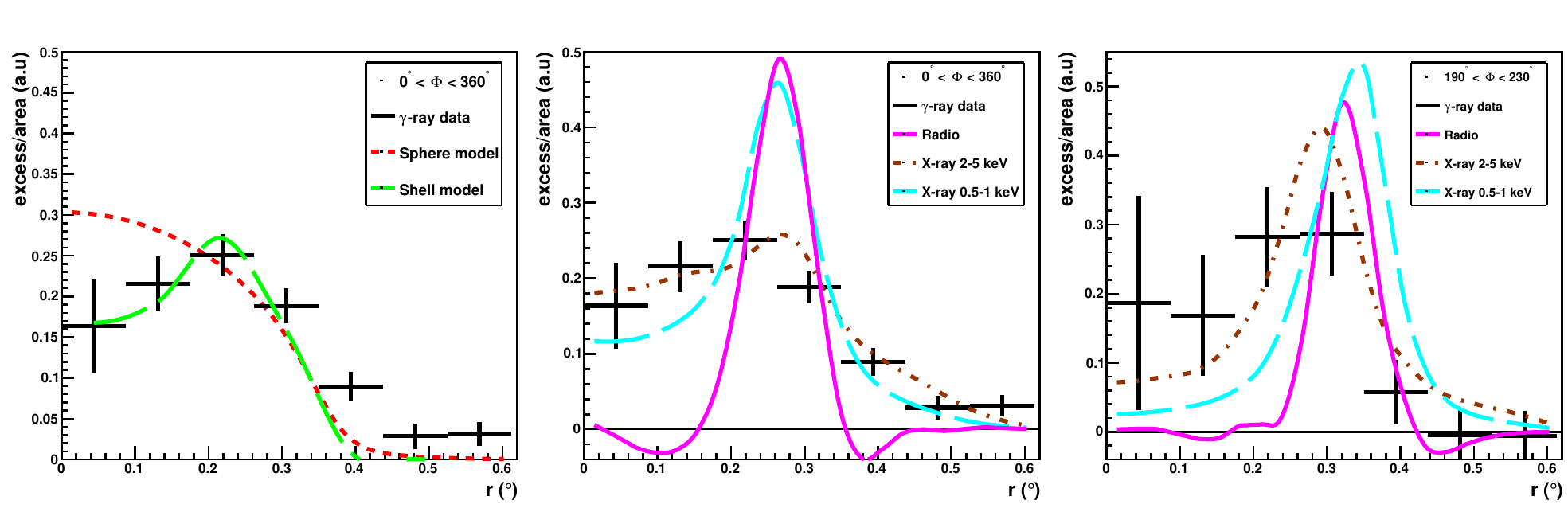} \caption{Radial profiles: The shell model's best-fit center point position (see Tabel \ref{table:1}) serves as a common center for the radial profiles of the $\gamma$-ray, X-ray, and radio data.
  {\it {In  left panel:}}
  The radial profile of the spatial distribution of the TeV $\gamma$-ray emission is shown as black crosses, which depict the measured TeV data points and their errors. The dashed red line corresponds to the radial profile of the sphere morphology and the green  line to the shell model.
  {\it {In center}} and {\it {right panel:}} Radial profiles in the TeV $\gamma$-ray, radio, 0.5 -- 1~keV, and 2 -- 5~keV X-ray regimes for different regions of RCW~86 are shown. While the center panel covers the entire position angle range, 
  the right panel only shows the SW region  with position angles between 
$190^{\circ}$ and $230^{\circ}$ 
(position angle $0^{\circ}$ corresponds to north and $90^{\circ}$ to east;
see Fig. \ref{fig::Skymap_rcw86}).  Black crosses are measured VHE excess points. The low-energy  (0.5 -- 1~keV, dashed cyan line) and high-energy ( 2 -- 5~keV, dotted brown line) X-ray band data  and the radio  data (solid magenta line) were smoothed with  the H.E.S.S.-PSF to account for the different angular resolution of the different instruments.  The radio data are from  Molonglo Observatory Synthesis Telescope \citep[MOST; see][]{whiteoak::1996} and the X-ray data  are from the {XMM-–\it{Newton}} X-ray telescope 
  \citep{broersen::2014}. The $\gamma$-ray and  X-ray data are normalized so that the area underneath the curve is equal to one; the radio data are scaled such that the area underneath the data points is equal to 0.5.
  }
\label{fig::RadialesProfile_rcw86}
\end{figure*}

To settle the question of whether the TeV $\gamma$-ray emission is shell-like as found in the optical, radio, and X-ray bands, a two-dimensional log-likelihood fit was applied to the uncorrelated sky maps taking  the morphological model and instrument response into account. 
To distinguish between different morphological structures, we used two alternative models: a sphere and a shell model. The models are obtained by line-of-sight integrals of a three-dimensional emitting sphere or shell. In the case of the shell model, the free parameters are the center position and the inner and outer radius. The sphere model is derived from the shell model by fixing the inner radius to zero.  Both models have uniform emissivity.
Table \ref{table:1} summarizes the fit results. The shell structure is favored over the sphere structure with a significance of about $8\, \sigma$.
In Fig. \ref{fig::RadialesProfile_rcw86} (left panel), the radial profile of the TeV  emission itself is given together with the fit results of both models.  


\begin{table*}
\caption{Results of the two-dimensional log-likelihood fit of morphological models to the H.E.S.S. data. ($\alpha_{center,J2000}$, $\delta_{center,J2000}$) gives the center of the sphere and shell models. The parameters $r_{in}$ and $r_{out}$ stand for the inner and outer radius of the sphere. In the case of the sphere, $r_{in}$ is kept at zero pc.}  
\label{table:1}  
\centering      
\begin{tabular}{|c|c|c|c|c|c|c|}
\hline 
Model & $\alpha_{center,J2000}$& $\delta_{center,J2000}$ & r$_{in}$ $[$pc$]$& r$_{out}$ $[$pc$]$ & Log(likelihood) \\ 
\hline 
sphere & $14^{\mathrm{h}}42^{\mathrm{m}}58^{\mathrm{s}} \pm 0^{\mathrm{h}}0^{\mathrm{m}}9^{\mathrm{s}} $
 & $-62\degr25\arcmin48\arcsec \pm 0\degr1\arcmin1\arcsec$ & -- & $15.9\pm 0.6$ &$-4093.35$\\ 
\hline 
shell &  $14^{\mathrm{h}}42^{\mathrm{m}}53^{\mathrm{s}}\pm  0^{\mathrm{h}}0^{\mathrm{m}}7^{\mathrm{s}}$ & $-62\degr25\arcmin48\arcsec \pm 0\degr1\arcmin48\arcsec$  & $9.5 \pm 1.4$ & $14.8 \pm 1.0$&$-4046.82$\\ 
\hline 
\end{tabular} 
\end{table*}

To facilitate the comparison, radial excess profiles of VHE $\gamma$-ray,   X-ray, and radio emission are presented in the center and right panels of Fig. \ref{fig::RadialesProfile_rcw86}. The X-ray data is split into two energy bands: 0.5 -- 1~keV and 2 -- 5~keV. The latter exhibits a higher amount of nonthermal emission in comparison to the low-energy band.  
The nonthermal emission in RCW~86 is not evenly distributed. \citet{broersen::2014} have shown that the NE region is dominated by nonthermal emission, while the SW region exhibits similar amounts of hot thermal and nonthermal emission. In the NW and SE regions, however, thermal emission prevails.
In Fig. \ref{fig::RadialesProfile_rcw86},   the central panel  depicts the radial profile of the whole SNR, whereas in the right panel only the data of the SW region within the position angle range{\footnote{Position angle $0^{\circ}$ corresponds to north and $90^{\circ}$ to east}} of  
$190^{\circ}$ to $230^{\circ}$ 
 were used to determine the profiles.   The radio data are from  Molonglo Observatory Synthesis Telescope  \citep[MOST;][]{whiteoak::1996} and the X-ray data are taken with the XMM--{\it{Newton}} X-ray telescope  \citep{broersen::2014}. 
The radial profile of the whole remnant (Fig. \ref{fig::RadialesProfile_rcw86}, center panel) clearly shows that the TeV and the X-ray emission of the energy range between 2 -- 5~keV are correlated, whereas in the case of low energetic X-ray emission (0.5 -- 1~keV) and radio emission a weaker luminosity is found in the central region and  the shell-like emission is more pronounced. 
 The radial profiles of the SW region of TeV $\gamma$-ray emission,  low- and  high-energy X-ray radiation (Fig. \ref{fig::RadialesProfile_rcw86}, right panel) again seem to reveal  a correlation between the 2 -- 5~keV X-ray emission and the TeV-emission. In addition, it seems that the maximum of the 2 -- 5~keV  X-ray emission is slightly more off-center than the maximum of the TeV emission, but this
is not significant because of the low statistics of the TeV data.  The maxima of the low-energy  X-ray (0.5 -- 1~keV) and  radio emission are at larger radii. In summary, the TeV emission and the high-energy X-ray emission (2 -- 5~keV) are correlated throughout the remnant, whereas radio and low-energy X-ray emission regions are further away from the center and show lower levels of emission in the central region of the remnant. 

To study the influence of the different physical conditions like shock speeds and densities on the $\gamma$-ray emission, spectra of two different subregions (SW and NE) were extracted.  Figure \ref{fig::Skymap_rcw86} shows the regions, their position, and their size. They are both centered around the best-fit position of the 3D-shell model. A Gaussian function was fitted to the radial profile of the $\gamma$-ray data and the $1\sigma$ interval around the mean of the fit was then taken as the minimal and maximal radii of the regions. In both regions, a detection significance of about 10~$\sigma$ has been achieved. The resulting spectrum is well described by a power-law model, which was fitted to the data up to an energy of 20~TeV. 
  It is not possible to distinguish between a simple power-law spectrum and one with an exponential cutoff because of the lower statistics of the quadrant data.
We obtained spectral indices of $\Gamma = (2.5 \pm 0.2_{stat} \pm 0.2_{sys}) $ and of $\Gamma = (2.2 \pm 0.1_{stat} \pm 0.2_{sys})$  as well as a differential flux normalization at 1~TeV of $\Phi_{0} =  (0.7 \pm 0.1_{stat} \pm  0.1_{sys}) \times 10^{-12}\mathrm{cm}^{-2}\mathrm{s}^{-1}\mathrm{TeV}^{-1}$ and $\Phi_{0} =  (0.7 \pm 0.1_{stat} \pm  0.1_{sys}) \times 10^{-12}\mathrm{cm}^{-2}\mathrm{s}^{-1}\mathrm{TeV}^{-1}$  in the NE and SW regions, respectively. Thus, there is no significant spectral difference between both regions.  The results are consistent with the spectrum of the whole remnant.

\section{Discussion}
\label{sec::model}

\begin{figure*}
\centering
\includegraphics[width=17cm]{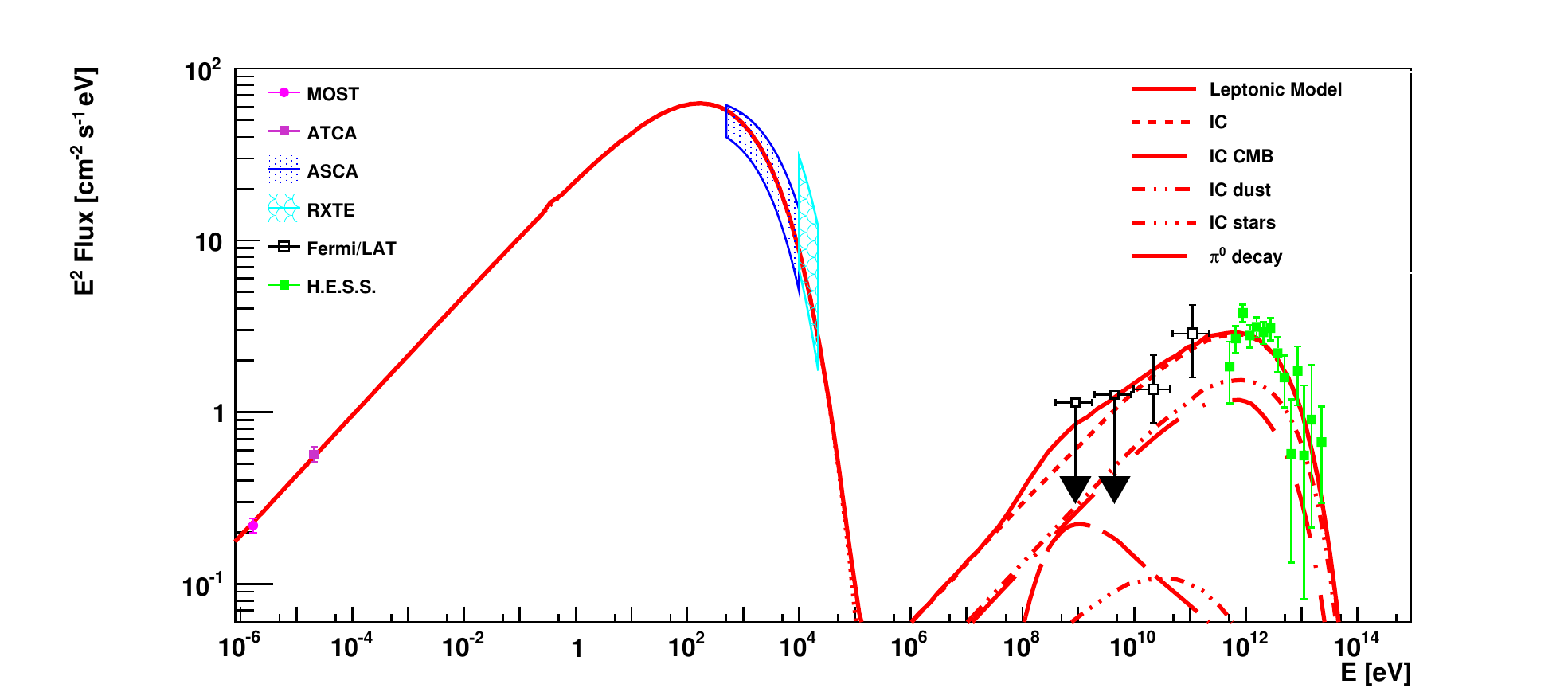} \caption{Spectral energy distribution of RCW 86 for a leptonic scenario. The red  solid lines denote the total broadband emission from the one-zone modeling discussed in Sect. \ref{sec::model} The dotted lines show the IC-contributions and the dashed line is that of the $\pi^0$-decay contribution.
      The radio data points are from Molonglo at 408 MHz and Parkes at 5 GHz \citep{caswell::1975,lemoine::2012}. 
      X-ray data are  from ASCA and RXTE from \citet{lemoine::2012}. The {\it{Fermi}}-LAT data points are taken from \citet{yuan::2014} and the H.E.S.S. data are from this analysis.}
\label{fig::Modellierung_SED_lep}
\end{figure*}

SNRs are thought to be the primary sources for the bulk of Galactic cosmic-ray protons  with energies up to $\sim 3\ \mathrm{PeV}$, but  the final proof is still lacking. 
The broadband, nonthermal emission of these sources is produced by accelerated particles through several channels, e.g., synchrotron, inverse Compton, nonthermal Bremsstrahlung, and neutral pion decay. To achieve insights into the $\gamma$-ray production taking place in SNRs, the different processes have to be disentangled and analyzed. 
We present the results of the broadband data of RCW~86 in two different scenarios for radiation mechanism for the $\gamma$-ray emission: inverse Compton (IC) scattering of electrons off ambient photons (leptonic scenario) or proton-proton interaction with the ambient medium (hadronic scenario). For both cases, we have modeled the broadband emission of RCW~86 with a simple static one-zone model \citep[presented in ][]{acero::2010}, where  
the radiation of the different wavebands is produced within the same region with a constant magnetic field. 
The current 
energy distributions of the particles (electrons and/or protons)  are given by a power law with an exponential cutoff.  Synchrotron radiation and IC scattering on the cosmic microwave background  and the  local interstellar (optical and infrared) radiation fields \citep[see][]{porter::2005} were taken into account with energy densities of 0.66~eV~cm$^{-3}$ (dust) and 0.94~eV~cm$^{-3}$ (stars). In  the case of the hadronic scenario, the $\gamma$-ray production was calculated following  \citet{kelner::2006}. Nonthermal Bremsstrahlung is neglected because of the low ambient density $\leq 1$~cm$^{-3}$. The modeling was done under the assumption that the distance of RCW~86 is 2500 pc, the outer radius is 15 pc, and the shell thickness is 5 pc.  The emission at all wavelengths was calculated for the shell region. The  stationary one-zone model is however an oversimplification because  the radio morphology and the nonthermal X-ray morphology differ, which cannot be explained by this kind of model.
The  spectral energy distribution presented in Fig. \ref{fig::Modellierung_SED_lep} is composed of the VHE $\gamma$-ray spectra presented in Sect. \ref{sec::Results},  the {\it{Fermi}}-LAT data points \citep{yuan::2014},  the X-ray spectra of ASCA and RXTE \citep{lemoine::2012}, and the radio data from the Molonglo at 408 MHz and Parkes at 5 GHz \citep{caswell::1975,lemoine::2012}.

In the leptonic case, the broadband data can be described by a one-zone model using an  electron spectrum  with a spectral index of $\Gamma_e \approx 2.3$, an exponential cutoff at $E_{cut} \approx 19$~TeV, and a magnetic field of $B \approx 22~\mu\mathrm{G}$ (see Fig. \ref{fig::Modellierung_SED_lep}). These results are comparable to those of  \citet{lemoine::2012} and  \citet{yuan::2014} and the magnetic field is comparable to estimates by \citet{vink::2006}.  The total kinetic energy of all electrons above 1~GeV amounts to $\sim 10^{48}\ \mathrm{erg}$, which is about 0.1\% of the total energy of a typical Type Ia supernova ($10^{51}\ \mathrm{erg}$).  We obtain an upper limit for the energy injected into accelerated protons of   $\sim10^{49} / n_{\mathrm{cm}^{-3}}\ \mathrm{erg}$ with the effective density $n_{\mathrm{cm}^{-3}} = n/ 1 \, \mathrm{cm}^{-3}$ also taking into account  protons
with the same spectral index as the electrons and a conservative chosen energy
cutoff of $100\ \mathrm{TeV}$. A higher amount of  energy  is not compatible with the {\it{Fermi}}-LAT upper limits (see Fig. \ref{fig::Modellierung_SED_lep}).  This energy limit implies an electron-to-proton ratio above $1\ \mathrm{GeV}$ of $K_{ep} \geq 10^{-1}$. 
These results are in good agreement with those obtained by \citet{lemoine::2012}.

\begin{figure*}
\centering
\includegraphics[width=17cm]{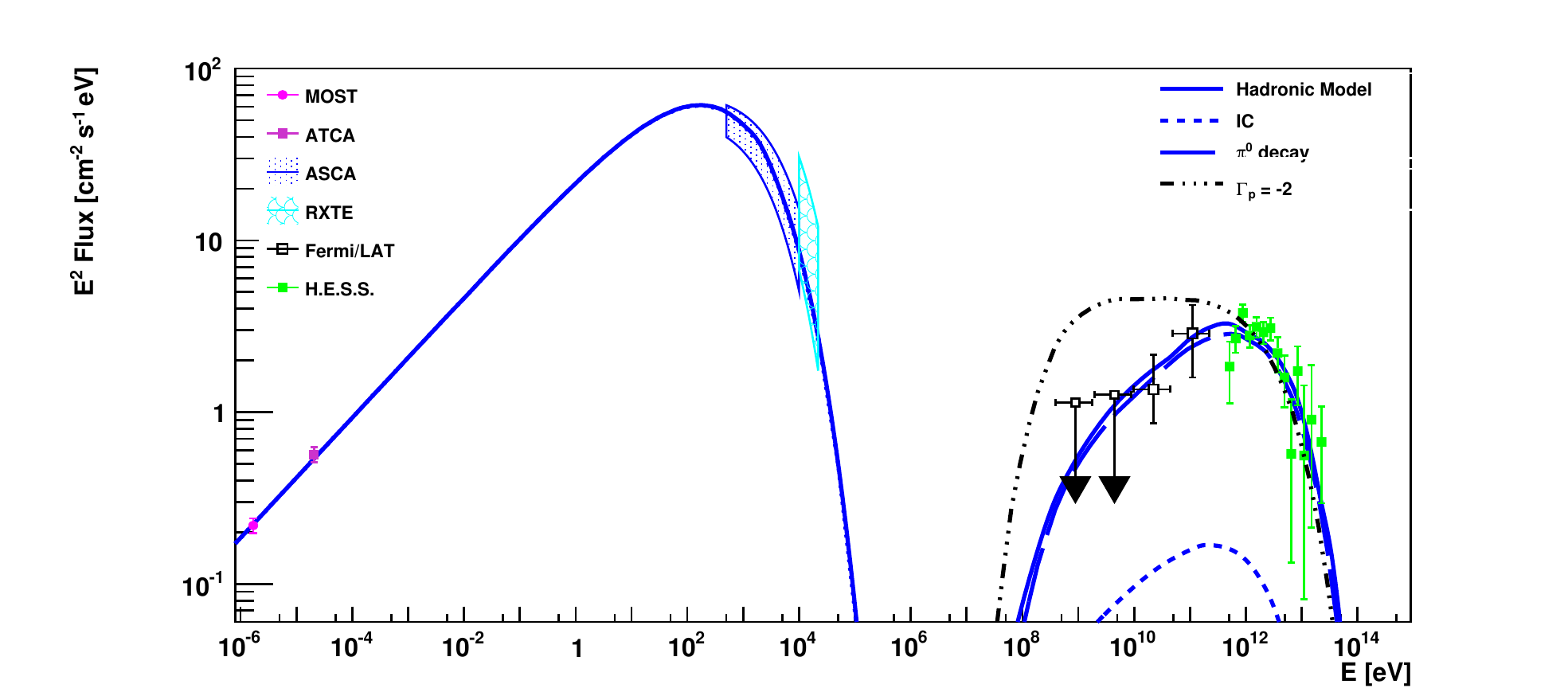} \caption{Spectral energy distribution of RCW 86 for a hadronic scenario. The   solid blue lines denote the total broadband emission from the one-zone modeling discussed in Sect. \ref{sec::model} The dotted lines show the IC-contributions and the dashed line is that of the $\pi^0$-decay contribution. The black dashed dotted line shows the results for a proton spectrum with $\Gamma_p = 2$.
      The radio data points are from Molonglo at 408 MHz and Parkes at 5 GHz \citep{caswell::1975,lemoine::2012}.
      X-ray data are  from ASCA and RXTE from \citet{lemoine::2012}. The {\it{Fermi}}-LAT data points are taken from \citet{yuan::2014} and the H.E.S.S. data are from this analysis.}
\label{fig::Modellierung_SED_had}
\end{figure*}

In the case of the hadronic scenario, the $\gamma$-rays are produced via proton-proton interactions and subsequent neutral pion decay whereas the X-rays are still produced via synchrotron radiation of high-energetic electrons. Therefore, the electron fraction has to be lower and the magnetic field stronger.
A standard proton spectrum with $\Gamma_p = 2$ and a total energy in accelerated protons at  emission time of $W_p \approx 2.1 \times 10^{50}/n_{\mathrm{cm}^{-3}} \ \mathrm{erg}$ would describe the H.E.S.S. measurements, but is incompatible with the {\it{Fermi}}-LAT results (see Fig. \ref{fig::Modellierung_SED_had} black dashed line). This was already pointed out by \citet{lemoine::2012}.
The proton spectrum, which reproduces the $\gamma$-ray data, has a spectral index of $\Gamma_p \approx 1.7$ and a cutoff energy of $E_{cut} = 35$~TeV. The index lies between the spectral index of the test particle approach and the spectral index with strong modified shocks \citep{malkov::1999,berezhko::1999}. The blue line in Fig.~\ref{fig::Modellierung_SED_had}  presents this hadronic model. 
The total energy of all protons above 1~GeV is $W_p \approx 9 \times 10^{49} / n_{\mathrm{cm}^{-3}} \ $~erg, which means that a fraction of $0.1/ n_{\mathrm{cm}^{-3}} $ of the supernova (Type Ia) energy has to be converted into high-energy protons.
The electron-to-proton ratio is $K_{ep} = 10^{-3}$. 
To model the radio and X-ray data, an electron spectrum with spectral index of  $\Gamma_e \approx 2.3$ and a cutoff energy of $E_{cut} = 9$~TeV is needed  and a magnetic field strength of 100~$\mu$G. This is comparable to two different estimates: one by \citet{voelk::2005} of $99^{+46}_{-26}\,\mu$G, which was deduced from the thickness of the filaments in the SW region, and another one calculated by \citet{Arbutina::2012} of a volume-averaged magnetic field of $\sim 70\ \mu$G. The total kinetic energy of the protons above 1~GeV was found to be $W_p \approx 9 \times 10^{49} /n_{\mathrm{cm}^{-3}} \ $~erg, which means that about 10\%$/ n_{\mathrm{cm}^{-3}} $ of the supernova (Type Ia) energy has to be converted into high-energy protons. 
\\
As already mentioned, the hadronic model cannot reproduce the radio data with an electron spectrum that has  the same spectral index as the proton spectrum ($\Gamma_p = 1.7$). 
This is in conflict  with the expectation that electrons and protons exhibit the same dynamics at relativistic energies up to an energy where electron synchrotron losses become important.

\citet{lemoine::2012} applied a two-zone model to the data to overcome this problem. They introduced a separate second leptonic population  to explain the radio emission with the consequence that the lower limit on the magnetic field was  $50~\mu$G.  The X-ray emission was reproduced with an injection spectrum with a power-law index of $\Gamma = 1.8$, a break at  3~TeV, caused by synchrotron cooling, and an exponential cut-off at 20~TeV. The energy injected into hadrons was $\sim 7 \times 10^{49} / n_{\mathrm{cm}^{-3}}\ \mathrm{erg}$. These results  fulfill the observational constraints,  but the problem remains that the proton spectrum is particularly hard.

When we compare the modeling results of both the leptonic and hadronic models, we find that neither can be ruled out. The hadronic model encounters problems by describing the data in a self-consistent way, i.e., it is not possible to describe the radio data with electron and proton spectra with the same spectral index, which would be expected for relativistic particles under the assumption of diffuse shock acceleration theory. Better statistics and possibly more detailed models are needed to solve this question.

Our study of the morphological correlation between radio, X-rays, and $\gamma$-rays has shown (see Sect. \ref{sec::Results}) that radio and the X-ray emission in the energy range between 0.5 and 1~keV do not coincide with the $\gamma$-ray emission, while the  emission in the X-ray regime 2 -- 5~keV shows a correlation with the TeV $\gamma$-ray emission.   This  X-ray emission (2 -- 5~keV) is spatially near to  a Fe-K-line.   \citet{rho::2002} argued that the hard X-ray continuum is synchrotron radiation produced by electrons, which are accelerated in the reverse shock. 
It is therefore possible that some of the VHE gamma-ray emission comes
from the region shocked by the reverse shock. A possible hint for this is provided by the more centrally filled morphology of the VHE
gamma-ray emission with respect to the radio emission, however, the sensitivity of the gamma-ray is not sufficient to draw any firm conclusions about this.

\section{Conclusion}
In  this work, we presented the first strong evidence that RCW~86 shows a shell-type morphology in TeV $\gamma-$rays. The TeV $\gamma-$ray spectrum favors an exponential cutoff power law with $\Gamma  = 1.59 \pm  0.22_{stat} \pm  0.2_{sys}$  and a cutoff energy $E_{cut} = 3.47 \pm  1.23_{stat} \pm  2.2_{sys}\ \mathrm{TeV}$ rather than a pure power law. 

The broadband SED can be well described by a simple leptonic one-zone model with a magnetic field strength of $B \approx 22\: \mu$G and a $\Gamma \approx 2.3$ power-law electron spectrum with an exponential cutoff at $E_{cut} \approx 19\ \mathrm{TeV}$. The kinetic energy of all electrons above $1\ \mathrm{GeV}$ is about 
0.1\% of the kinetic energy of the supernova explosion of $10^{51}\ \mathrm{erg}$. 

Modeling the broadband data using a hadronic one-zone model requires a hard proton spectrum with index $\Gamma_p \approx 1.7.$ This model is incompatible with a conventional $\propto E^{-2}$ acceleration spectrum, but lies between the spectral index value of the test particle approach and the one with strong modified shocks \citep{malkov::1999,berezhko::1999}. The total energy of all protons above  1~GeV  is  $W_p \approx 9 \times 10^{49} /n_{\mathrm{cm}^{-3}}$~erg,  when  a distance of 2.5 kpc are assumed. As a result of limited statistics neither the leptonic nor the hadronic model can be ruled out.

\begin{acknowledgements}
The support of the Namibian authorities and of the University
of Namibia in facilitating the construction and operation of H.E.S.S. is gratefully
acknowledged, as is the support by the German Ministry for Education and
Research (BMBF), the Max Planck Society, the German Research Foundation
(DFG), the French Ministry for Research, the CNRS-IN2P3 and the Astroparticle
Interdisciplinary Programme of the CNRS, the U.K. Science and Technology
Facilities Council (STFC), the IPNP of the Charles University, the Czech
Science Foundation, the Polish Ministry of Science and Higher Education, the South African Department of Science and Technology and National Research Foundation, and by the University of Namibia. We appreciate the excellent work of the technical support staff in Berlin, Durham, Hamburg, Heidelberg, Palaiseau, Paris, Saclay, and in Namibia in the construction and operation of the equipment.
The MOST is operated by The University of Sydney with support from the Australian Research Council and the Science Foundation for Physics within The University of Sydney.

\end{acknowledgements}

\bibliographystyle{aa} 
\bibliography{RCW} 

\end{document}